\documentclass[twocolumn,showpacs,preprintnumbers,amsmath,amssymb]{revtex4}
\usepackage{graphicx}% Include figure files
\usepackage{dcolumn}% Align table columns on decimal point
\usepackage{amscd}
\usepackage{bm}% bold math

\usepackage[mathscr]{eucal}
\date{\today}
\begin{document}

%\preprint{APS/123-QED}

\title[Gauge Group and Topology Change]{Gauge Group and Topology Change}
\author{Izumi Tanaka  }
 \email{tanaka-i@eng.setsunan.ac.jp}

\author{Seiji Nagami}%
 \email{nagami@atf.setsunan.ac.jp}


\affiliation{%
Education Center, Faculty of Engineering, Setsunan University, Neyagawa 572-8508, Japan
}

\begin{abstract}
The purpose of this study is to examine the effect of topology change in the initial universe.
In this study, the concept of $G$-cobordism is introduced to argue about the topology change of the manifold on which a transformation group acts. This $G$-manifold has a fiber bundle structure if the group action is free and is related to 
the spacetime in Kaluza-Klein theory or Einstein-Yang-Mills system. 
Our results revealed that  
fundamental processes of compactification in $G$-manifolds.
In these processes, the initial high symmetry and multidimensional universe
changes to present universe by the mechanism which 
lowers the dimensions and symmetries.

\end{abstract}

\pacs{04.20.Gz, 04.50.+h, 04.60.Gw.}% PACS, the Physics and Astronomy
%\date{today}                              
\maketitle

%%%%%%%%%%%%%%%%%%%%%%%
\newtheorem{thm}{Theorem}
\newtheorem{prop}[thm]{Proposition}
\newtheorem{cor}[thm]{Corollary}
\newtheorem{df}[thm]{Definition}
\newtheorem{lem}[thm]{Lemma}
\newtheorem{con}[thm]{Convention}
%%%%%%%%%%%%%%%%%%%%%%%
%\newcommand{\scri}{\Im}
%\def\scri{{\cal I}
%\newcommand\scri{\mathscr{I}}
%\newcommand\scri{\mathcal{I}}
%%%%%%%%%%%%%%%%%%%%%%%%%
%%%%%%%%%%%%%%%%%%%%%%%
%%%%%%%%%%%%%%%%%%%%%%%

\section{Introduction}
Various studies have been conducted on evolution of initial universe.
From the theory of unification, initial universe has high symmetry, multidimensional origin and 
 would not distinguish between visible and extra dimensions. 
For consistency with experience, 
topological separation of visible and extra dimensions by means of a certain kind of spontaneous symmetry breaking is needed, and it is required to give the differentiation of  the initial dimension into %a scale between 
extra and visible dimensions %(spontaneous compactification) 
\cite{Chodos-Detweiler1980}.
As a result, present topology would be product of compact manifold 
with our universe after the topological separation and the differentiation had succeeded. 
It is considered that such topological separation take place in the Plank era,
at which spacetime would be governed by quantum gravity.
The topological separation is specifically expressed by compactification as following:
%By a process of compactification, we mean a topological separation in which 
the initial spacelike hypersurface is $M_{\text{initial}}$ and the final hypersurface is $M_{\text{visible}}\times M_{\text{extra}}$. 
In this process, by a compactified universe, we mean a universe with spacelike 
topology $M_{\text{visible}}\times M_{\text{extra}}$.

In this paper we address the influence of the topology change to the spacetime with symmetry. 
Specifically, we study the compactification of multidimensional universe with high symmetry by the quantum gravity effect.
Kaluza-Klein (KK) space-time or Einstein-Yang-Mills system (EYMS) can be 
referred to as a spacetime relevant to such symmetry. 
The universe accompanied by a gauge field like these are described by $G$-manifold.
Studying about the influence of the topology change to such spacetime has a physical significance.

There is another viewpoint of introducing $G$-manifold as follows:
Under the standard model, there is only a restriction of a dimension for the multidimensional universe \cite{Witten}.
Since the general solutions of a higher dimensional Einstein equation do not have symmetry at all, it is possible to use a $G$-manifold to confer symmetry on them by additional conditions.
Introducing the symmetry as both uniformity and isotropy 
by means of $G$-manifold, quantum evolution of the multidimensional universe %using Wheeler-Dewitt equation 
has been analysed 
\cite{Bertolami-Mourao1991,Bertolami-Kubyshin-Mourao-1992,Bertolami-Fonseca-Moniz1997}.

At least in certain scope, it seems reasonable to think that quantum gravity may be conceived as a 
path integral over spacetimes \cite{Hawking1978}.  
One considers all cobordism joining the given initial and final spacelike hypersurfaces. 
Then it also seems reasonable to extend the cobordism to $G$-cobordisms 
having the
boundaries as hypersurfaces before and after the topology change 
when spacetimes have gauge symmetry.
%\cite{Bertolami-Mourao1991,Bertolami-Kubyshin-Mourao-1992,Bertolami-Fonseca-Moniz1997}.

We consider it significant to analyse the initial universe using a $G$-manifold from these facts.
Previous studies of topology change based upon cobordism theory 
have been argued, such as
cobordism \cite{Geroch1967,Tipler1985,Sorkin1986}, spin-Lorentz cobordism
 \cite{Gibbons-Hawking,Chamblin}, the Morse theory \cite{Alty1994,Ioniciou1997,Dowker-Garcia1998,Borde-Dowker-Garcia-Sorkin-Surya1999,Dowker-Garcia-Surya2000}, and surgery theory \cite{Dowker-Surya1998, Hartnoll2003}, up to now.  

When a non-degenerate Lorentz metric and a causal partial order on 
spacetime 
are not  satisfied, spatial topology change is not allowed within the classical  theory of relativity \cite{Geroch1967}. Also when weak energy condition is taken into consideration with the Einstein equation, similar result can be derived \cite{Tipler1985}.
Previously, topology change has been discussed by abandoning causal order and allowing closed timelike curves (CTCs) \cite{Sorkin1986}.

On the other hand, by abandoning CTC and giving priority to causal order, the causally continuous almost Lorentzian (CCAL) cobordism can be obtained \cite{Dowker-Garcia1998}.
Furthermore, it is possible to obtain causal continuous cobordism by using surgery \cite{Dowker-Surya1998, Hartnoll2003}.
Such CCAL cobordism has been introduced as a criterion of whether cobordism  contributes to the Sum Over Histories (SOH) or not.

An explosion of papers argue about the quantum evolution of universe can be described 
by SOH.
In Euclidean path integral \cite{Hartle-Hawking1983}, gravitational action $S_E$ is 
unbounded from below. Then the assertion that the contribution from a saddle point comprises the principal part of a wave function becomes meaningless. 
The convergence is achieved only by integrating along a complex contour in the the space of complex metrices. 
There are a number of  inequivalent contours along which the path integral converges, 
each dominated by different saddle-points with  different forms for the 
wave functions \cite{Halliwell-Louko1989,Halliwell-Hartle1990}.
On the other hand, there is room for argument in a Lorentzian path integral 
for the improvement of outgoing mode proposal and singularity \cite{Vilenkin}.
Thus, the Euclidean and the Lorentzian path integrals have problems in their formulation.

In Causal Dynamical Triangulations (CDT) which also discusses path integrals,  causality serves as an critical point in their discussion 
\cite{Loll-Westra03, Loll-Westra-Zohren06, Ambjorn-Loll-Westra-Zohren07}.
Since the universe superposed in the Euclidean quantum gravity is inherently unstable 
simply because it results in too many configurations contributing to the path integral, 
the Lorentzian spacetime is assembled with encoded causality 
\cite{Ambjorn-Gorlich-Jurkiewicz-Loll2008}.

However, previous studies cannot capture the concrete effect of topology change 
resulting in manifold with symmetry.  
We do not discuss from a viewpoint of wave function.
Instead we advance an argument by paying attention to the specific path corresponding to topology change of the multidimensional universe with gauge symmetry.

We would like to clarify anew our position of dealing with topology change. 
Our approach is to utilize the $G$-cobordism theory:

First, $G$-manifolds with fiber bundle structure like Kaluza-Klein theory, Einstein-Maxwell system and Einstein-Yang-Mills system are set as the objects of topology change.
$G$ shows the transformation group (gauge group) which acts freely on a manifold.

Second, $G$-cobordism is introduced in order to deal with topology change of $G$-manifold.
$G$-cobordism is reduced into cobordism \cite{Geroch1967,Tipler1985,Sorkin1986}  when gauge group $G=\{1\}$. 

Third, when treating topology changes, causality should be respected.
Let CCAL be a criterion whether the cobordism is physically allowed or not.

CCAL metric of spacetime is degenerate at a finite number of isolated 
singularities. It is thought that finiteness of  quantum field propagation on the spacetime
  \cite{Dowker-Surya1998, Anserson-DeWitt1986, Harris-Dray1990}  
is related to causality continuum \cite{Dowker-Garcia1998, 
Borde-Dowker-Garcia-Sorkin-Surya1999, 
Dowker-Surya1998}. 
Topology changing solutions of the first order form of general relativity also deals with degenerate metric \cite{Horowitz1991,Borde1994}.

This paper discusses 
the topology change accompanied by higher to lower gauge
symmetry transition. 
In Section II, $G$-cobordism is surveyed and the basic theorem of $G$-cobordism %with a compact Lie group $G$ 
is proved.
In Section III, the concrete example of a $G$-cobordant manifold is given. 
In Section IV, conclusions and discussion of topology change in spacetime are made.

%%%%%%%%%%%%%%%%%%%%%%
\section{ Mathematical preliminaries}
 %%%%%%%%%%%%%%%%%%%%%
We specify mathematically the type of spacetime that is targeted.
Einstein-Maxwell system (EMS) and 
Einstein-Yang-Mills system (EYMS) have principal fiber bundle structure with gauge group, 
$G=U(1)$ for EMS 
and $G=SU(2)$ for EYMS, respectively. 
We use EGS to refer the system having principal fiber bundle structure with gauge group
$G$  such as EMS or EYMS. These systems  can be identified with $G$-manifold: 
\begin{thm}
Let $G$ be a compact Lie group and $\widetilde{M}$ be $G$-manifold.

If $G$ action is free, orbital space $\widetilde{M}/G$ will serve as a manifold, and  $\pi: \widetilde{M}\mapsto \widetilde{M}/G$ becomes projection map of 
$C^{\infty}$ principal fiber bundle where $G$ is the fiber.

The differentiation structure on $\widetilde{M}/G$ is unique and fulfills the following conditions:

(i) $\pi: \widetilde{M}\mapsto \widetilde{M}/G$ is $C^{\infty}$ map.

(ii) $h:\widetilde{M}/G \mapsto N$ is $C^{\infty}$ map if and only if $h \circ \pi$ is $C^{\infty}$ map.

\end{thm}

For Kaluza-Klein theory (KK) 
where gauge field is reduced from a higher dimensional $G$-manifold, 
the following relation is known.

\begin{thm}

Let pseudo Riemann manifold $(\widetilde{M}, \tilde{g} )$ be invariant under
transformation group $G$'s free action. Then $(\widetilde{M}, \tilde{g} )$ shows one-to-one correspondence to pseudo Riemann metric $g$ on orbital space $M=\widetilde{M}/G$, gauge field $A$, and scaler field $\gamma$ which takes value in inner product of Lie algebra $\mathfrak{g}$, such that
\begin{align}
A^{\prime} & = ad(a)A+daa^{-1}, a\in G\nonumber\\
\gamma^{\prime} & = \gamma\circ\text{ad(a)}.\nonumber
\end{align}

\end{thm} 
In order to deal with topology change of these $G$-manifolds, 
it is necessary to introduce the concept of 
$G$-cobordism which is extension of cobordism.
In our discussion, topology change is limited to $G$-cobordantness among manifolds.
We outline the $G$-cobordism theory for $G$-manifolds \cite{Conner-Floyd,Uchida_CBRD}.

Let $H$ be a closed Lie subgroup of $G$.
We denote by $(H)$ the set of Lie subgroups of $G$ that are conjugate to $H$.
\begin{df}
A family $\mathfrak{F}$ of closed subgroups of $G$ is called admissible
if  $(K)\subset \mathfrak{F}$ whenever 
$K\in \mathfrak{F}$.
If the isotropy subgroup of any point of $\widetilde{M}$ belongs to $\mathfrak{F}$,
then $\widetilde{M}$ is called $\mathfrak{F}$ free $G$ manifold.
Moreover, suppose that $\widetilde{M}$ has a boundary $\partial \widetilde{M}$
and that $\mathfrak{F}$ has a admissible subfamily $\mathfrak{F}^{'}$.
Then $\widetilde{M}$ is called $(\mathfrak{F},\mathfrak{F}^{'})$ free 
if $\widetilde{M}$ is $\mathfrak{F}$ free and $\partial \widetilde{M}$ is $\mathfrak{F}^{'}$ free.
\end{df}

\begin{df}
Let $\widetilde{M}$ be a $(\mathfrak{F}, \mathfrak{F'})$ free
$n$-dimensional oriented compact $G$-manifold. 
Then $\widetilde{M}$ is called null
$G$-cobordant if there exist an $(n+1)$-dimensional $\mathfrak{F}$ free compact
$G$-manifold $W$ and an $n$-dimensional compact $G$-invariant submanifold $\widetilde{M}_{1}$ of $\partial W$, such that

(1) $\widetilde{M}$ is $G$-isomorphic to $\widetilde{M}_{1}$,

(2) $\partial W - \widetilde{M}_{1}$ is $\mathfrak{F}$ free.

This is expressed as $\widetilde{M} \sim0$.

Further, two $n$-dimensional $(\mathfrak{F}, \mathfrak{F'})$ free oriented compact
$G$-manifolds $(\widetilde{M}, G)$, $(\widetilde{M}^{\prime}, G)$ are called
$G$-cobordant if the direct sum as a set $(\widetilde{M}, G)+(-\widetilde{M}^{\prime}, G)$
is null $G$-cobordant.
\end{df}

This relation is expressed as $(\widetilde{M}, G)\sim(\widetilde{M}^{\prime}, G)$.

The relation $G$-cobordant is equivalent relation among the set of 
$(\mathfrak{F}, \mathfrak{F'})$ free $n$-dimensional compact $G$-manifolds.

\begin{con}
We denote by $[\widetilde{M},G]$ the $G$-cobordant class of
a $(\mathfrak{F}, \mathfrak{F'})$ 
free oriented compact $G$-manifold $(\widetilde{M}, G )$, and by
$\Omega_{n} (G; \mathfrak{F}, \mathfrak{F'})$  the  set of
$G$-cobordant classes.
\end{con}

  Note that $\Omega_{n}
(G:\mathfrak{F}_{1}) $ ( $\mathfrak{F}_1=\{\{1\}\}$,  $\{1\}$: trivial subgroup) is the set of $G$-cobordant classes of
$G$-manifolds  where compact Lie group $G$ acts on freely.

\begin{df}
 Let $g:N \mapsto Y$ be a singular $n$-manifold of
 a topological space $Y$.
For  $g:N\mapsto Y$,
 a partition of some natural number $I=(k_{1}, k_{2}, \cdots, k_{m})$,
and cohomology classes $x\in H^{*}(Y; \mathbb{Z}_{2}), y\in H^{*} (Y; \mathbb{Z})$,

set
\begin{align}
w_{I, x}(g) & =\langle w_{k_{1}}(N) \cdots w_{k_{m}}(N) (g^{*}x)
\hspace{0.5mm}, [N]_{2} \rangle\in\mathbb{Z}_{2}\nonumber\\
p_{I, y}(g)  & =\langle p_{k_{1}}(N) \cdots p_{k_{m}}(N) (g^{*}y)
\hspace{0.5mm}, [N] \rangle\in\mathbb{Z}.\nonumber
\end{align}
These are called bordism Stiefel-Whitney number and bordism Pontrjagin number of
$g$  for $I$ respectively .

If $f_{G}$ is the classifying map (see Appendix \ref{PBCS}) $f_{G}: \widetilde{M}/G\mapsto \mathbb{B}G$
 of a free $G$-action $\psi : \widetilde{M}\times G\mapsto \widetilde{M}$, then
 $w_{I,x}(f_{G})$ and $p_{I,y}(f_{G})$ are denoted by
 $w_{I,x}(\widetilde{M}, G)$ and $p_{I,y}(\widetilde{M}, G)$, and are called
 the bordism Stiefel-Whitney number and bordism Pontrjagin number 
 of $\psi $
 respectively.
\end{df}

The following facts are known \cite{Conner-Floyd,Uchida_CBRD}.
\begin{thm} \label{SWP}
Let $g:N \mapsto Y$ be a singular $n$-manifold of
 a topological space $Y$.
 Suppose that $[N,g]=0\in \Omega _{n}(Y).$
 Then all of the
 bordism Stiefel-Whitney numbers and bordism Pontrjagin numbers of $g$
 vanish.
\end{thm}

\begin{thm} \label{G-cob}
Let $\widetilde{M}$ be an $n$-dimensional free $G$-manifold without
boundary.
 Then 
$\left[ \widetilde{M}, G\right] =0\in \Omega_{n}
(G:\mathfrak{F}_{1}) $ as the element of $G$-cobordism group if and only if all of the bordism Pontrjagin numbers and the bordism
Stiefel-Whitney numbers of the $G$-manifold vanish.
\end{thm}

\begin{thm} \label{module}

Defining $p_{*}([\widetilde{M}, G])=[\widetilde{M}/G, f_G]$,
 we have the  following isomorphism of degree -dim$(G)$ as graded $\Omega_{*}$-module,

$p_{*} :\Omega_{*}(G )\mapsto\Omega_{*}(\mathbb{B}G), $%\nonumber
%\end{equation}
\hspace{2mm}where $\Omega_*(G) =\oplus_{n\ge 0} \Omega_{n} (G) $.
\end{thm} 

Since $\mathbb{B}G$ is infinite CW complex,
we need the following well known theorem about CW complexes \cite{Spanier}.

\begin{thm} ( Cellular approximation theorem)\label{cellular}

Let $X$ be an $n$-dimensional CW complex
and $f:X\mapsto Y$  a continuous map from $X$ to an another CW complex $Y$. 
Then there exists a cellular map $f^{'}:X\mapsto Y^{(n)}$
that is homotopic to $f$,
where $Y^{(n)}$ denote the $n$-skelton of $Y$.
\end{thm} 
Combining the above theorems, we can prove the following:
\begin{thm} \label{G-cobordism}
Let $\widetilde{M}$  be an $n$-dimensional oriented compact
$G$-manifold without boundary.
Suppose that the homology group $H_{*} (\mathbb{B}G)$ has no torsion and that the Thom
homomorphism $\mu: \Omega_*(\mathbb{B}G)\mapsto H_{*}(\mathbb{B}G)$ is  surjective.
 Then
$\left[ \widetilde{M}, G\right] =0\in\Omega_{n} (G:\mathfrak{F}%
_{1}) $ 
 if and only if all of the bordism Pontrjagin numbers and the bordism
Stiefel-Whitney number of the $G$-manifold vanish.

{\it Proof.}

First suppose that  $\left[  \widetilde{M} , G\right]  =0 \in\Omega_{n} (G)$.
Let $p_{*} (\left[  \widetilde{M} , G\right])=\left[  \widetilde{M}/G, f_G  \right]$

Then from Theorem \ref{SWP}, all of the bordism Pontrjagin numbers and the bordism Stiefel
Whitney numbers of classifying map $f_G: \widetilde{M}/G \mapsto \mathbb{B}G $ are zero.

Thus all of the bordism Pontrjagin numbers and the bordism Stiefel Whitney
numbers of the $G$-manifold are zero.

Conversely,
suppose that all of the bordism Pontrjagin numbers and the bordism Stiefel Whitney
numbers of the $G$-manifold are zero.

This means that all of the bordism Pontrjagin numbers and the bordism Stiefel
Whitney numbers of the classifying map
 $  f_G: \widetilde{M}/G \mapsto \mathbb{B}G $ are zero.

By the Theorem \ref{cellular} (cellular approximation theorem), we have a map $f^{'}_G: \widetilde{M}/G\mapsto (\mathbb{B}G)^{(n+1)}$
that is homotopic to $f$ in $\mathbb{B}G$, where $(X)^{k}$ denote the 
$k$-skeleton of $X$.
Thus $\left[  \widetilde{M}/G, f_G  \right]  =0 \in \Omega_{n-d}(\mathbb{B}G)$ by Theorem \ref{G-cob}.
This implies that $[\widetilde{M}/G, f_G]=[\widetilde{M}/G, f^{'}_G]=0 \in \Omega _{n-d}(\mathbb{B}G)$.
Furthermore,  from Theorem \ref{module}, $p_{*} :\Omega_{m}(G)\mapsto\Omega_{m-d}(\mathbb{B}G)$ is an
isomorphism as $\Omega$-module.
Therefore, since  $p_{*} ([\widetilde{M},G])=\left[  \widetilde{M}/G, f_G \right]
$, we have that $\left[  \widetilde{M},G \right]  =0$. 
\hspace{50mm} $\Box$

\end{thm} 
Then condition of topology change can be summarized as follows:
\lq \lq In the topology change based on $G$-cobordism, the bordism Stiefel-Whitney number $w_{I,x}(\widetilde{M},G)$ and bordism Pontrjagin number $p_{I,y}(\widetilde{M},G)$ are equal before and after the topology change.\rq\rq \\
$G$-cobordism is reduced into cobordism when $G=\{1\}$.

\section{Topology change}\label{TC}
We specify the type of topology change that is considered. 
In order for given compactification process $M_{\text{initial}} \rightarrow M_{\text{visible}}\times M_{\text{extra}}$ possible, one needs to prove the 
existence of a CCAL cobordism $W$ such that $\partial W= M_{\text{initial}} \sqcup M_{\text{visible}}\times M_{\text{extra}}$ \cite{Hartnoll2003}.   
Our argument extends cobordism which deals with compactification to $G$-cobordism.

The extension to G-cobordism of compactification is given as follows:
From now on, let $G$, $G_1$, and $G_2$ be  compact Lie groups such that 
they admit orientations
that are preserved  by both left and right translations,
and that the skeletons of their classifying spaces are finite.
Let $\widetilde{M}$ be an oriented compact differentiable
$n$-dimensional $G$-manifold without boundary.
Note that the quotient map
$\widetilde{M} \mapsto \widetilde{M}/G$ is a principal $G$-bundle.
We denote its  classifying map by $f_G:\widetilde{M}/G \mapsto \mathbb{B}G$.
Suppose that the topology of a $G$-manifold $\widetilde{M}$ changes to a direct
product of two G-spaces with dim $G_{i}\geq 1$ for $i=1,2$($G=G_1\times G_2$) : 
\begin{equation}
(\widetilde{M},G) \longrightarrow (\widetilde{M}_{1},G_{1}) \times (\widetilde{M}_{2},G_{2})
\end{equation}
%If this occurs, we say that $(\widetilde{M},G)$ splits to lower dimension.
This means, they are $G$-cobordant:
\begin{equation}
\left[  \widetilde{M}, G \right]  \sim \left[  \widetilde{M}_{1}, G_{1} \right] \times \left[ \widetilde{M}_{2}, G_{2}\right] .
\end{equation}

In KK interpretation, topology change $\left[ \widetilde{M}, G \right]\sim \left[ \widetilde{M}_1,  G_1 \right]\times\left[ \widetilde{M}_2,  G_2 \right]$ with interpolating
manifold $\widetilde{W}$ shows that $\widetilde{M}$,  $\widetilde{M}_1$, 
and $\widetilde{M_2}$ are spatial hypersurface, and $\widetilde{W}$ is 
whole spacetime with 
$\partial \widetilde{W}= \widetilde{M}\sqcup \widetilde{M}_1\times 
\widetilde{M_2}$. 
Further,  $\widetilde{M}/G$ corresponds to 
visible dimensions interacting with 
gauge field originated in gauge group $G$.

In EGS interpretation, 
there is a cobordantness between $M=\widetilde{M}/G$ and  $M_1(=\widetilde{M}_1/G_1)\times M_2(=\widetilde{M}_2/G_2)$:
$\partial W =M\sqcup M_1\times M_2$, where
$W=\widetilde{W}/G$ is whole spacetime interacting with 
gauge field originated in gauge group $G$. 
This is because $\partial\big(\widetilde{W}/G\big)=(\partial\widetilde{W})/G$ is
 satisfied by $G$'s free action. 
Above all base spaces  are oriented, since $G$ preserves the orientation.
The existence of the manifolds $\widetilde{W}$ and $W$ make given topology changes possible.

Based on this premise, two examples are discussed below.

\textbf{Example 1}
The situation that spatial hypersurface with higher symmetry changes to direct product of hypersurfaces with lower symmetry is considered.

Let $\xi(P,q,X,{\bf Spin}(4))$ be a principal $\mathbf{Spin}(4)$-bundle.
 Let $p_{i} 
:\mathbf{Spin}(4) =\mathbf{SU}(2)\times\mathbf{SU}(2)\to\mathbf{SU}(2)$ 
denote the projection
map to the i-th factor. Then we obtain $\mathbf{SU}(2)-$bundle $P_{i}%
=P\times_{p_{i}}\mathbf{SU}(2)$. Let $c_{2}^{(i)}(\xi)\in H^{4}(X;\mathbb{Z})$ denote the
second Chern class of $P_{i}\to X$.
\begin{prop}\label{A}
Let $\xi(\widetilde{M},p,M,{\bf Spin}(4))$ be a principal $\mathbf{Spin }(4)$-bundle over an
oriented closed 4-manifold with the projection map $p:\widetilde{M}\to M$. 
Then
the oriented $\mathbf{Spin}(4)$-bordism class 
$\left[\widetilde{M}, \mathbf{Spin}
(4)\right]$ splits to lower dimension if and only if 
$\sigma(M)=0$ 
and $c_{2}
^{(i)}(\xi)=0\in H^{4}(M)$ for $i$=1,2,
where $\sigma (M)$ denotes the signature of 
$M$
(see Appendix \ref{S4n}).

\textit{Proof.}

Suppose that $\left[\widetilde{M}, \mathbf{Spin}(4)\right]$ splits to $\left[\widetilde{N}
_{1},\mathbf{SU}(2)\right]\times\left[\widetilde{N}_{2},\mathbf{SU}(2)\right]$:

$\left[\widetilde{M}, \mathbf{Spin}(4)\right]\sim \left[\widetilde{N}
_{1},\mathbf{SU}(2)\right]\times\left[\widetilde{N}_{2},\mathbf{SU}(2)\right]$.
 Then the
$\mathbf{SU}(2)$-bundle $\widetilde{N}_{1}\to N_{1}$ is  principal bundle
over 1,2, or 3-dimensional manifold $N_{1}$. Thus $w_{1}(N_{1})=w_{2}
(N_{1})=p_{1}(N_{1})=0$. 
Moreover, for the classifying map $\alpha_{1}
:N_{1}\to\mathbb{B}\mathbf{SU}(2)$, 
the induced homomorphism $\alpha^{*}_{1}:H^{*}(\mathbf{SU}(2);\mathbb{Z})\to
H^{*}(N_{1})$ is trivial, 
since $H^{*}(\mathbb{B}\mathbf{SU}(2))$ is generated by
$c_{2}\in H^{4}(\mathbf{SU}(2);\mathbb{Z})$.
 Thus $\left[\widetilde{N}_{1},\mathbf{SU}(2)\right]=0$.  
Similarly,
$\left[\widetilde{N}_{2},\mathbf{SU}(2)\right]=0$.  
Thus we have only to study the
necessary and sufficient condition for $\left[\widetilde{M}, \mathbf{Spin}(4) \right]=0$.

Let $\xi(\widetilde{M},p,M,{\bf Spin}(4))$ denote the principal $\mathbf{Spin}(4)$ bundle
with classifying map $f=f_{1}\times f_{2}:M\to\mathbb{B}\mathbf{SU}(2)\times\mathbb{B}\mathbf{SU}
(2)$.
 Suppose that $\left[\widetilde{M}, \mathbf{Spin}(4)\right]=0$. 
Since $M$ bords, $\sigma(M)=0$. 
Moreover, $c_{2}^{(1)}(\xi)=f^{*}_1c_{2}=0$.
 Similarly,
$c_{2}^{(2)}(\xi)=f^{*}_2c_{2}=0$.

Conversely, suppose that $\sigma(M)=0$ and $c_{2}^{(i)}(\xi)=0 (i=1,2)$.
 Then we have
that $p_{1}(M)=3\sigma(M)=0$, and that 
$w_{2}^{2}(M)\equiv p_{1}(M)=0$ modulo
2. 
Moreover, we have that $f^{*}x=0$ for all $x(\neq0)\in H^{*}(\mathbb{B}\mathbf{Spin}(4))$, since
$c_{2}^{(1)}(\xi)=c_{2}^{(2)}(\xi)=0.$ Thus $\left[\widetilde{M}, \mathbf{Spin}(4)\right]=0$. \\
This completes the proof.
\hspace{30mm} 
 $\Box$
\end{prop}

If we restrict the conditions of above proposition, 
we have similar results for higher dimension.

The first case is given as follows:

(i)$\left[\widetilde{M}, \mathbf{Spin}(4)\right]$ splits to $\left[\widetilde{N}
_{1},\mathbf{SU}(2)\right]\times\left[\widetilde{N}_{2},\mathbf{SU}(2)\right]$ with 
2-dimensional manifold $N_{1}=\widetilde{N}_{1}/\mathbf{SU}(2)$ 
and
3-dimensional manifold $N_{2}=\widetilde{N}_{2}/\mathbf{SU}(2)$ 

\begin{cor}\label{A-1}

Let $\xi(\widetilde{M},p,M,{\bf Spin}(4))$ be a principal $\mathbf{Spin }$(4)-bundle over an
oriented closed 5-manifold with the projection map $p:\widetilde{M}\to M$. 
Then
the oriented $\mathbf{Spin}(4)$-bordism class 
$\left[\widetilde{M}, \mathbf{Spin}
(4)\right]$ splits to lower dimension if and only if 
$\omega_2(M)\omega_3(M)=0$.

\textit{Proof.}

This corollary can be proved similarly as Proposition \ref{A}.
\hspace{70mm} 
$\Box$
\end{cor}

The second case is given as follows:

(ii)$\left[\widetilde{M}, \mathbf{Spin}(4)\right]$ splits to $\left[\widetilde{N}
_{1},\mathbf{SU}(2)\right]\times\left[\widetilde{N}_{2},\mathbf{SU}(2)\right]$ with 

3-dimensional manifolds $N_{1}=\widetilde{N}_{1}/\mathbf{SU}(2)$ and 
$N_{2}=\widetilde{N}_{2}/\mathbf{SU}(2)$.

\begin{cor}\label{A-2}

Let $\xi(\widetilde{M},p,M,{\bf Spin}(4))$ be a principal $\mathbf{Spin }$(4)-bundle over an
oriented closed 6-manifold with the projection map $p:\widetilde{M}\to M$. 
Then
the oriented $\mathbf{Spin}(4)$-bordism class 
$\left[\widetilde{M}, \mathbf{Spin}
(4)\right]$ splits to lower dimension if and only if 
$ \omega_2^3(M)=\omega_3^2(M)=0$
%, $c_{2}^{(i)}(\xi)=0\in H^{4}(M)$ 
and $ \big(c_{2}^{(i)}(\xi)\big)_2\omega_2(M)=0 $ 
% \in H^{6}(M)
for $i$=1,2,  
where $\big(c_2^{(i)}(\xi) \big)_2$ denotes the reduction of 
$c_2^{(i)}(\xi)$ modulo 2.

\textit{Proof.}

This corollary can be proved similarly as Proposition \ref{A}.
\hspace{70mm} $\Box$

\end{cor}
{\bf Remark:} The similar argument is possible also about $\mathbf{SU}(3)\times \mathbf{SU}(3)$ instead of  $\mathbf{Spin}(4)$.
Furthermore, $G_1\times G_2$, a group with higher symmetry is possible. \\

\textbf{Example 2}
The situation that spatial hypersurface with higher symmetry changes to  hypersurface with lower symmetry is considered.

Let $\xi(\widetilde{M},p,M,G_1\times G_2)$ be a principal $G_1\times G_2$
bundle with the classifying map $f=f_1\times f_2:M\mapsto \mathbb{B}G_1\times \mathbb{B}G_2$.
Suppose that $H^{*}(\mathbb{B}G_2;\mathbb{Z})$ is generated by only one element
$l\in H^{d}(\mathbb{B}G_2;\mathbb{Z})$.
($H^{*}(\mathbb{B}G_2;\mathbb{Z}_2)$ is generated by only one element
$l_2\in H^{d}(\mathbb{B}G_2;\mathbb{Z})$)
Set $\eta =f_2^{*}l\in H^{d}(M;\mathbb{Z})$ and $l_2\in H^{d}(\mathbb{B}G_2;\mathbb{Z}_2)$.\\

\begin{prop}\label{B}
$\left[  \widetilde{M}, G_1\times G_2 \right]  $ topologically changes to $\left[
\widetilde{M}_{1}, G_1\right]  \times\left[ G_2, G_2 \right]  $ if and only if

$p_{I,x\otimes l^k}(\widetilde{M},G_1\times G_2)=
w_{I,x_2\otimes (l_2)^k}(\widetilde{M},G_1\times G_2)=0$ hold
for all triples $(x,I,k)$ of cohomology classes $x\in H^{*}(\mathbb{B}G_1;\mathbb{Z})$,
 $x_2\in H^{*}(\mathbb{B}G_1;\mathbb{Z}_2)$,
 partition of natural numbers $I$,
and   natural numbers  $k\in \mathbb{N}=\{1,2,\ldots \}$.
In particular, if $\eta =0$,
then $[\widetilde{M},G_1\times G_2]$ splits to lower dimension.

\textit{Proof.}

First suppose that 
$\left[\widetilde{M},G_1\times G_2 \right]\sim \left[\widetilde{M}_1,G_1\right]\times \left[G_2,G_2\right]$.
In this case, we obtain the  principal $G_1\times G_2$ bundle 
$\widetilde{M}_1\times G_2\mapsto M_1=\widetilde{M}_1/G_1$
with the classifying map $g=g_1\times pt:M_1\mapsto \mathbb{B}G_1\times \mathbb{B}G_2$,
where $pt$ denote a constant map.
Then, for all $I$ and $x\otimes y \in H^{i}(\mathbb{B}G_1;\mathbb{Z})\otimes H^{j}(\mathbb{B}G_2;\mathbb{Z})$ with 
$j\ge 1$,
we obtained $p_{I,x\otimes y}(\widetilde{M}, G_1\times G_2)=
\langle p_{I}(M)f^{*}(x\otimes y),[M]\rangle=
\langle p_{I}(M_1)g^{*}(x\otimes y),[M_1]\rangle=\langle p_{I}(M_1)g_1^{*}x(pt)^{*}y,[M_1]\rangle=0$.

Thus $p_{I,x\otimes l^{k}}(\widetilde{M}, G_1\times G_2)
=0$ holds for all
$x$, $I$, and $k$.
Similarly we have $w_{I,x_2\otimes (l_2)^{k}}(\widetilde{M}, G_1\times G_2)
=0$.

Conversely, suppose that \\
$p_{I,x\otimes l^{k}}(\widetilde{M})=
\langle p_{I}(M)f_1^{*}(x) \eta ^{k},[M]\rangle=0$
hold for all $x$, $I$, and $k$.
Let $\widetilde{M}_1=f_1^{*}\mathbb{E}G_1\mapsto M$ denote the 
$G_1$ bundle induced by $f_1:M\mapsto \mathbb{B}G_1$.
Then, by applying  the direct product with  $G_2$ bundle $G_2\to c$, we obtain a $G_1\times G_2$ bundle 
$\widetilde{M}_1\times G_2\mapsto M\times c=M$
with the classifying map $f_1\times pt:M_1\mapsto \mathbb{B}G_1\times \mathbb{B}G_2$,
where $c$ denotes one point.
Thus for all $x\otimes y \in H^{i}(\mathbb{B}G_1;\mathbb{Z})\otimes H^{j}(\mathbb{B}G_2;\mathbb{Z})$
with $j\geq 1$,
we have that 
$\langle p_{I}(M)f^{*}(x\otimes y),[M]\rangle=0=\langle p_{I}(M)(f_1\times pt)^{*}(x\otimes y),[M] \rangle$. 

Moreover, by identifying  $H^0(\mathbb{B}G_2;\mathbb{Z})$ with $\mathbb{Z}$,
we have that 
$\langle p_{I}(M)f^{*}(x\otimes y),[M]>=
y\langle p_{I}(M)f_1^{*}x,[M]\rangle=
\langle p_{I}(M)(f_1\times pt)^{*}(x\otimes y),[M] \rangle $
for $x\otimes y \in H^{i}(\mathbb{B}G;\mathbb{Z})\otimes H^{0}(\mathbb{B}G_2;\mathbb{Z})$.
We also have similar equality for the bordism Stiefel-Whitney numbers.
Therefore by virtue of Theorem \ref{G-cobordism}, we have that
$\left[\widetilde{M}, G_1\times G_2\right]\sim \left[\widetilde{M}_1, G_1\right]\times \left[G_2,G_2\right]$.
This completes the proof. \hspace{20 mm}$\Box$
\end{prop}
{\bf Remark:}  Note that $H^{*}(\mathbb{B}\mathbf{U}(1);\mathbb{Z})$
is generated by the first Chern class 
$c_1\in H^{2}(\mathbb{B}\mathbf{U}(1);\mathbb{Z})$ and $\mathbf{U}(1)\cong S^1$.
$H^{*}(\mathbb{B}\mathbf{SU}(2);\mathbb{Z})$ is generated by the second Chern class 
$c_2\in H^{*}(\mathbb{B}\mathbf{SU}(2);\mathbb{Z})$ and 
$\mathbf{SU}(2)\cong S^3$.
See Appendix A.\\

\section{Conclusions and discussion}
%%%%%%%%%%%%%%%%%%%%%%%%%%%%%%%%%%%%%%%%%%%%%%%%%%
In this paper, we have explored the topology change among $G$-manifolds.
We first proved the theorem specifying condition for the cobordantness among 
$G$-manifolds.  
Based upon this theorem,  we described two examples 
%Consider now the implication of the examples 
in the \S \ref{TC}.
%It is considered that 
The interpolating manifold (cobordism) is thought to be related to the spacetime 
from a viewpoint that spatial topology change emerges in spacetime.
Results of topology change in $G$-manifold can be translated into 
those in KK and EGS. 
Moreover, it is possible to introduce the concept of extra dimensions (plus visible dimensions) in EGS 
 , not to mention KK.

Examples 1 and 2 show the fundamental processes of compactification in $G$-manifolds.
From these processes, the initial high symmetry and multidimensional universe
changes to present universe by the mechanism which 
lowers the dimensions and symmetries.

Here, we explain the essential characteristic
of these two processes by EGS interpretation .
In example 1, the hypersurface splits to two hypersurfaces, and their gauge symmetries are lowered.
In example 2, the hypersurface lowers its gauge symmetry and changes its topology 
without changing its dimension.
In both example 1 and 2, topology changes are accompanied by higher to lower gauge 
symmetry transitions.

After the compactified universe emerges by the process of example 1, following secondary process can be considered: 
it is possible to interpret hypersurface in example 2 as visible (or extra) dimension. 
In this case, interpolating manifold of spacetime consists of product of trivial cobordism of extra (or visible) dimension and cobordism of visible (or extra) dimension.
In this processes, as clear from the Proposition \ref{B}, 
conditions are imposed only to visible (or extra) dimension before topology change: 

$p_{I,x\otimes l^k}(\widetilde{M},G_1\times G_2)=
w_{I,x_2\otimes (l_2)^k}(\widetilde{M},G_1\times G_2)=0$, 

\noindent where $\widetilde{M}$ is visible (or extra) dimension.
As a result, change of topology and gauge symmetry emerge only in visible (or extra)
dimension.

When the above process is applied to the extra dimension, 
significant effect related to superstring theory can occur. 
Since there is no restriction for geometrical configuration of 
extra dimension after topology change, the number of generations of fundamental fermions related to this 
configuration can be changed \cite{Candelas}.   
From the above-mentioned processes and their combination, geometrical configuration connected to the differentiation is given.

Each example is explained in the following.
In example 1, we show the necessary and sufficient condition that $\left[  \widetilde{M}, \mathbf{Spin}(4) \right]  $ and $\left[
\widetilde{N}_{1}, \mathbf{SU}(2) \right]  \times\left[ \widetilde{N}_{2}, \mathbf{SU}(2) \right]  $ are  G-cobordant 
 in Proposition \ref{A}, Corollary \ref{A-1} and Corollary \ref{A-2}. 
Interpolating manifold is denoted by $\widetilde{W}$.

(A) KK interpretation:

$(6+a)$-dimensional null cobordant hypersurface $\widetilde{M}$ can topology 
change to direct product of $(3+b)$-dimensional null cobordant  hypersurface $\widetilde{N}_1$ and $(3+a-b)$-dimensional null cobordant hypersurface $\widetilde{N}_2$. 
($(a, b)=(4, 1)$, $(4, 2)$, $(4, 3)$, $(5, 2)$, $(5, 3)$, $(6, 3)$) 
It is possible to think either  $\widetilde{N}_1$ or 
$\widetilde{N}_2$ to correspond to extra dimensions.

(B) EGS interpretation:

(B-i)
Base spaces are cobordant:
for $W=\widetilde{W}/\mathbf{Spin}(4)$, $M=\widetilde{M}/\mathbf{Spin}(4)$,
 and $N_i=\widetilde{N}_i/\mathbf{SU}(2)$, ($i=1, 2$), 
there exists a relation $\partial W= (M \sqcup N_{1} \times N_2)$.
Under $W$ being whole spacetime, spatial hypersurface $M$ are compactified  to $N_1\times N_2$ by topology change.
$M$, $N$ and $N_1$ are null cobordant. 
It is possible to think either $N_1$ or 
$N_2$ to correspond to extra dimensions.

(B-ii) 
Corresponding to (A-i)
we have dim$M=a$, dim$N_1=b$ and dim$N_2=a-b$.
Since $W$ corresponds to spacetime, $M$ is higher dimensional  hypersurface,  $N_1$ is $b$ dimensional hypersurface, and $N_2$ is $a-b$ dimensional  hypersurface.

%%%%%%%%%%%%%%%%%%%%%%%%%%%%%%%%%%%%%%%%%%%%%%%%%
Let us switch our attention to example 2. 
In example 2, we show the necessary and sufficient condition for $\left[  \widetilde{M}, G_1\times G_2 \right]  $ and $\left[
\widetilde{M}_{1}, G_1\right]  \times\left[ \widetilde{M}_2(=G_2), G_2 \right]  $ to be G-cobordant in  Proposition \ref{B}. 
Interpolating manifold is denoted by $\widetilde{W}$.

(A') KK interpretation:

If dim$\widetilde{M}=r$, dim$G_1=s$ and dim$G_2=t$, $r$-dimensional null cobordant hypersurface $\widetilde{M}$ can topology 
change to direct product of $(r-t)$-dimensional %null cobordant
 hypersurface $\widetilde{M}_1$ and $t$-dimensional %null cobordant
 hypersurface $\widetilde{M}_2$. 
 It is possible to think that  
$\widetilde{M}_2$ corresponds to extra dimensions. 
$\widetilde{M}_2/G_2$ becomes a point.

(B') EGS interpretation:

(B'-i) Corresponding to (A'), there can be 
a topology change between closed manifolds $M$ and $M_1$
with dim$M=$dim$M_1=r-s-t$. (dim$M_2=0$)

(B'-ii) If $r-s-t$ is even, dimension of whole spacetime cobordism $W$ becomes odd. 
If $\widetilde{M}/G_1\times G_2$ is homeomorphic to $\widetilde{M}_1 $, for example if we take 
$\widetilde{M}=\widetilde{Q}_1\times G_2$ where 
$\widetilde{Q}_1 $ is homeomorphic to $\widetilde{M}_1 $, then,  
$\chi(M)=\chi(M_1)$ which shows this Lorentzian cobordism \cite{Sorkin1986}.
In this case, there can be no CTC.

(B'-iii) Under (B'-ii), if $W$ admits trivial second Stiefel-Whitney class $\omega_2(W)=0$,  $\partial W=M \sqcup M_{1} $ becomes spin-Lorentz cobordism \cite{Chamblin}.  

(B'-iv) 
If $\widetilde{M}/G_1\times G_2 $ is diffeomorphic to $\widetilde{M}_1/G_1 $, 
there is a possibility that there is no topology change among spatial hypersurfaces and W is a Lorentzian cobordism \cite{Geroch1967}. 
For example, this is realized as $\widetilde{M}=\widetilde{R}_1\times G_2$ where  
$\widetilde{R}_1$ is diffeomorphic to $\widetilde{M}_1 $.

(B'-v)  Let $M$ and $M_1$ be obtained from asymptotically flat, non-compact space $U$ and $U_1$ with one-point compactification, respectively. Further, let  $\widetilde{M}$ and $\widetilde{M}_{1}$ be G-cobordant, where  
$M=\widetilde{M}/G$ and $M_1=\widetilde{M}_{1}/G$.  
Then it is apparent that base spaces $M$ and $M_{1}$ are cobordant. $U$ and $U_1$ are also cobordant 
based on the discussion by Dowker and Garcia \cite{Dowker-Garcia1998}.
It is necessary to examine whether the cobordism which interpolates two hypersurfaces observe
the causality or not.

The case where the dimension of $G$-cobordism is five or more is considered.
CCAL cobordism can be obtained if the cobordism $\widetilde{W}$ can be made simply connected by equivariant surgery
 ($G$-surgery) and admitted to have the different handle decomposition with no $1$-handles and $(N-1)$-handles (dim $\widetilde{W}=N$) \cite{TGM}.
This is similar argument as Hartnoll \cite{Hartnoll2003}.
Above discussion can be applied to the KK interpretations (A) and (A').

About EGS interpretation (B), following arguments will be attained. 
If $G$-cobordism $\widetilde{W}$ can be made simply connected by equivariant surgery and different handle decomposition is chosen for eliminating the unnecessary handles
in base space, CCAL cobordism can be obtained in the spacetime.
Further, from the homotopy perfect sequence of fiber bundle 
$\pi_1 (G) \rightarrow \pi_1(\widetilde{W}) \rightarrow \pi_1(W) \rightarrow  \pi_0(G)$, 
the spacetime being simply connected  is guaranteed.

Finally, we have the following comments:
for both the requirement of extra dimension from the theory and 
the observation of the spacetime dimension to be consistent, 
the theory must contain certain mechanism as compactification.
It should be also taken into consideration that a series of gauge symmetry lowering  occurred in the universe.
From this viewpoint, it is reasoned that 
dynamical compactification should occur,  
i.e., our geometrical mechanism emerges as dynamical mechanism.  

In all of these topology changes is that 
the process is accompanied by higher to lower gauge
symmetry transition.
This process is different from Higgs mechanism.
This is because the groups act on direct product spaces, and 
the degree of freedom to lower symmetry group is not used for generation of the boson.

 It should be added that the cobordantness among spacetimes themselves can be discussed  instead of spatial hypersurfaces.

%\section*{Acknowledgments}

%This section should come before the References. Dedications and funding
%information may also be included here.

\appendix

%%%%%%%%%%      %%%%%%%%%%%%%%%%%%%
%%%%%%%%%%    %%%%%%%%%%%%%%%%%%%
%%%%%%%%%%  %%%%%%%%%%%%%%%%%%%
\section{Principal bundles and Classifying Spaces \cite{EDM} }\label{PBCS}

  Let $G$ be a compact Lie group.
  A principal $G$-bundle $\xi (P,p,X,G)$ consists of total space $P$,
  base space $X$, and projection map $p:P\mapsto X$
  such that $G$ acts $P$  transitively on the right,
  and that the orbit $\tilde{x}G$ of $\tilde{x}\in P$
  is exactly the fiber $p^{-1}(x)$ of $x=p(\tilde{x})$.
  
  The universal $G$-bundle $\pi :\mathbb{E}G\mapsto \mathbb{B}G$
  is the principal $G$-bundle such that $\pi _{i}(\mathbb{ E}G)=\left\{1\right\}$
  for all $i\in \mathbb{ N}\cup\{0\}$.
  $\mathbb{B}G$ is called the classifying space of $G$.
  It is known that $\mathbb{B}G$ is determined uniquely up to 
  weak homotopy equivalences.
   The following theorem is a basic fact for the theory of principal bundles.   
\begin{thm} (Classifying Theorem) : 
   The set of the isomorphism classes $[G;X]$ of principal $G$-bundles 
over $X$ has one to one correspondence
with the set of homotopy classes $[X;\mathbb{B}G]$ of continuous maps from 
$X$ to $\mathbb{B}G$.  
\end{thm}
Throughout this paper,
we have assumed that $H^{*}(\mathbb{B}G;\mathbb{Z})$ is torsion free.
Related to the torsion elements of $H^{*}(\mathbb{B}G;\mathbb{Z})$,
the following theorems are known. 
\begin{thm}If $H_{*}(G;\mathbb{Z})$ has no p-torsion, then $H_{*}(\mathbb{B}G;\mathbb{Z})$ has no p-torsion.
\end{thm}
\begin{thm}
If $H_*(\mathbb{B}G;\mathbb{Z})$ has no torsion, then $H_*(G;\mathbb{Z})$ has no torsion and  is a polynomial ring generated by even dimensional element.
\end{thm}
%%%%%%%%%%********************************************************************
\begin{align}
H^{*}(\mathbb{B}\mathbf{U}(n);\mathbb{Z}) & =H^{*}(\mathbb{B}\mathbf{GL}(n,\mathbb{C});\mathbb{Z})=\mathbb{Z}[c_{1}, c_{2}, \cdots,
c_{n}]\nonumber\\
H^{*}(\mathbb{B}\mathbf{SU}(n);\mathbb{Z}) & =H^{*}(\mathbb{B}\mathbf{SL}(n,\mathbb{C});\mathbb{Z})=\mathbb{Z}[c_{2}, \cdots,
c_{n}] \nonumber
%H^{*}(\mathbb{B}\mathbf{Sp}(n);\mathbb{Z}) & =\mathbb{Z}[q_{1}, q_{2}, \cdots, q_{n}]\nonumber\\
%H^{*}(\mathbb{B}\mathbf{O}(n):K_{2}) & =H^{*}(\mathbb{B}\mathbf{GL}(n,\mathbb{R}); \mathbb{Z}_{2})=K_{2}[w_{1},
% \cdots, w_{n}]\nonumber\\
%H^{*}(\mathbb{B}\mathbf{SO}(n):K_{2}) & =H^{*}(\mathbb{B}\mathbf{SL}(n,\mathbb{R}); %\mathbb{Z}_{2})=K_{2}[w_{2},
%\cdots, w_{n}]\nonumber\\
%H^{*}(\mathbb{B}\mathbf{SO}(2m+1):K) & =K[p_{1},p_{2}, \cdots, p_{m}]\nonumber\\
%H^{*}(\mathbb{B}\mathbf{SO}(2m):K) & =K[p_{1},p_{2}, \cdots, p_{m-1}, \chi]\nonumber
\end{align}
where %$K_{2}$ is field of characteristic 2, 
$K$ is the field which is
not characteristic 2, $c_{i}$ is $i$th Chern class, 
%$q_{i}$ is $i$th simplectic
%Pontrjagin class, $w_{i}$ is $i$th Stiefel-Whitney class, $p_{i}$ is $i$th
%Pontrjagin class, $\chi$ is Euler class, 
and deg$c_{i}=2i$.
%, deg$q_{i}%
%=$deg$p_{i}=4i$, deg$w_{i}=i$, deg$\chi=2m$.

\section{Signature of a 4n-manifold}\label{S4n}

Let $M$ be a closed oriented manifold of dimension $4n$.
Then we have a non-degenerate bilinear form
$I:H_{2n}(M;\mathbb{R})\times H_{2n}(M;\mathbb{R})\mapsto \mathbb{R}$
defined by $I(x,y)=\langle x\cup y,[M]\rangle$,
which is called the intersection form of $M$.
Then $H_{2n}(M;\mathbb{R})$ splits to the direct sum 
$H_{2n}(M;\mathbb{R})=H_{+}\oplus H_{-}$,
where $I|H_{+}$ is positive definite and
$I|H_{-}$ is negative definite. 
Set $\sigma (M)=dim _{\mathbb{R}}(H_{+})-dim_{\mathbb{R}}(H_{-})$.
We call $\sigma (M)$ the signature of $M$.
It is known that $\sigma (M)$ is a cobordism invariant of $M$ \cite {Milner}.

%\section*{References}

%References are to be listed in the order cited in the text in Arabic 
%numerals within square brackets. They can be 
%referred to indirectly, e.g.~``$\ldots$
%in the statement \cite{beeson}.'' or used directly, 
%e.g.~``$\ldots$ see [2] for examples.'' List references 
%using the style shown in the following examples. For journal names,
%use the standard abbreviations.  Typeset references in 9 pt roman.

\end{document}